\documentclass{elsart}

\begin{document}
\begin{frontmatter}
\title{Traversal-Time Distribution for a Classical Time-Modulated
Barrier}

\author{Jos\'e L. Mateos\thanksref{EMAIL}}
\address{Instituto de F\'{\i}sica, 
Universidad Nacional Aut\'onoma de M\'exico, \\
Apartado Postal 20-364, 01000 M\'exico, D.F., M\'exico}
\thanks[EMAIL]{E-mail: mateos@fenix.ifisicacu.unam.mx; \\
Fax: (525) 622-5015; Phone: (525) 622-5130}

\begin{abstract}
The classical problem of a time-modulated barrier,
inspired by the B\"uttiker and Landauer model to
study the tunneling times, is analyzed. We show that the
traversal-time distribution of an ensemble of non-interacting
particles that arrives at the oscillating barrier,
obeys a distribution with a power-law tail.

\end{abstract}
\end{frontmatter}


\section{Introduction}
The problem of obtaining the time involved in the tunneling process
in quantum mechanics is still a controversial issue, despite 
considerable efforts in recent years \cite{Reviews}.
In particular, in order to address this issue,
some authors have analyzed the tunneling through time-modulated
potential barriers \cite{Bula82,Bula85,Soko88,Garcia90,Leav91,Leav93,Pimp91,Azbel,AzMa93,LanMar93,Stov93,MarSa95,Wag95}.
One of the pioneer works in this area is the model introduced by
B\"uttiker and Landauer in 1982 \cite{Bula82} in which they consider the
transmission through a time-modulated rectangular barrier, and
introduced a characteristic time for the process. 
However, in the above-mentioned papers, there is practically no
mention of the corresponding classical problem;
although the classical limit is straightforward when the potential
barrier does not depend on time,
it is far from trivial when the potential is time modulated. 

In this paper I study the classical problem of a rectangular
time-modulated potential barrier, in order to analyze in detail the
traversal time distribution for an ensemble of classical particles.
This classical model was inspired, in part, by the B\"uttiker-Landauer
model mention above.

I will study first the case of a potential barrier located inside
a rigid box \cite{Mateos}. In this case, the classical orbits can
be periodic, quasiperiodic or even chaotic, depending on the parameters and
the initial conditions of the motion. In order to study the dynamics, I
derive first an area-preserving map that allow us to find the orbits for
all times. Then, I study the scattering problem of an ensemble of 
particles that interact with an oscillating rectangular potential barrier.  
In this case, I will show that the traversal time strongly 
depends on the arrival time of the incident particles.

There is a basic difference between these two problems:
(1) In the first case, what we have is the bounded problem of 
an oscillating barrier inside a rigid box of finite size. This means 
that an incident particle interact with the barrier not once but an 
arbitrary number of times, since the particle can cross the barrier 
region and then, after bouncing elastically in the box, returns to
the oscillating barrier. Then, the dynamics can
become chaotic, since we have the main ingredients: on one hand, 
sensitive dependence on initial condition or arrival times due to
the oscillating barrier, and on the other hand, bounded motion due to
presence of the finite box. 
(2) In the second case, we have a scattering problem in which
an incident particle interacts with the barrier only once. Of course, if
this is the case, the problem is straightforward and there is only a
single traversal time. But, if we consider an ensemble of $N$ 
noninteracting particles with slightly different initial conditions, 
say different initial velocities, then we can expect, in general,
$N$ different traversal times that can exhibit a complex distribution
of traversal times.

An approach to the problem of tunnelling times, that is closely related 
to the classical trayectories discussed here, is the Bohm
trajectory point of view \cite{Holl}. This approach has been used by 
Leavens and Aers \cite{Leav95} to give an unambiguous prescription for
calculating traversal times that are conceptually meaningful within that
interpretation. In particular, Leavens and Aers \cite{Leav91,Leav93} 
have treated in detail the case of a time-modulated rectangular barrier,
using Bohm's trajectory interpretation of quantum mechanics \cite{Holl}.
They calculate, among other things, transmission time distributions,
the transmission probability as a function of frequency and Bohm
trajectories.

\section{The model and the map}

Let us study the classical dynamics of a particle in a one-dimensional box,
inside of which there is an oscillating rectangular potential 
barrier \cite{Mateos}. This problem consists of a particle moving in one dimension under the action of a time-dependent potential $V(x,t)$. Since 
the Hamiltonian of this system is time dependent, the total energy of the particle is not conserved. The Hamiltonian is given by 
$H(x,p,t)=p^2/2m + V(x,t)$, where 
\begin{equation}
V(x,t)=V_0(x)+V_1(x)f(t).
\end{equation}
The potential $V_0(x)$ goes to infinity when $x<0$ or $x>l+b+L$, is equal to
the constant value $V_0$ when $l\le x\le l+b$, and otherwise is equal to
zero. Thus, what we have is an infinite potential well with a rectangular
potential barrier of width $b$ inside, as shown in Fig. 1a. This potential separates the box in
three regions: region I, $0\le x<l$ of width $l$; region II, where the
rectangular barrier is located, $l\leq x\le l+b$ of width $b$; and region
III, $l+b<x\le l+b+L$ of width $L$.

Clearly, the motion of a particle under the influence of the potential
$V_0(x)$ is regular, that is, we have periodic orbits and the energy is
conserved. However, if we add a time-dependent potential, we can obtain
periodic, quasiperiodic and chaotic orbits, as we will show below. The
potential $V_1(x)$ in eq. (1) is different from zero only inside the
interval $l\le x \le l+b$, where it takes the constant value $V_1$. The
function $f(t)$ in eq. (1) is assumed periodic with period $\tau$, that
is, $f(t+\tau) = f(t)$. In this way, as shown in Fig. 1a, what we have is an
oscillating potential barrier, with an amplitude which oscillates between 
$V_0 - V_1$ and $V_0 + V_1$, with frequency $\omega/2\pi$ and period 
$\tau=2\pi/\omega$. We will take $V_0>V_1$.

Let us now derive a map that describes the dynamics of a particle under this
potential. The motion is as follows: at the fixed walls at $x=0$ and $x=l+b+L
$, the particle bounces elastically, changing the sign of the velocity but
with the same absolute value. The other two points where there is a change
in the velocity is at the borders of the potential barrier at $x=l$ and 
$x=l+b$. The rest of the time the velocity is constant. Thus, 
the particle can gain or loose kinetic energy at $x=l$ and $x=l+b$. 
The phase space for a typical orbit is depicted in Fig. 1b.

We can analyze the dynamics using a discrete map from the time $t_n$ when
the particle hits the wall at $x=0$, until the next time $t_{n+1}$ when it
hits this wall again. Let us denote by $v_n$ the velocity of the particle
immediately after the $n-th$ kick with the fixed wall at $x=0$, and by 
$E_n$ the corresponding total energy. Clearly, $E_n=m{v_n}^2/2$. After
traveling the distance $l$, it arrives at the left side of the barrier
after a time of flight $l/v_n$, where a change in the velocity occurs. 
To determine this change let us consider the following: In region I,
the total energy of the
particle is given by $E_n=m{v_n}^2/2$ which is just the kinetic energy,
because in this region the potential energy is zero; when the particle
enters region II, the kinetic energy $E_n^{\prime }$ is changed to 
$E_n-V_0-V_1f(t_n+l/v_n)$, that is, the total energy minus the value of the
potential energy at the time of arrival $t_n+l/v_n$. If we denote the new
velocity by $v_n^{\prime }$ (see Fig. 1b), then 
$E_n^{\prime }=m{v_n^{\prime }}^2/2$ and
we obtain in this way the change in energy as: 
\begin{equation}
E_n^{\prime }=E_n-V_0-V_1f\biggl(t_n+{\frac l{v_n}}\biggr).
\end{equation}
Clearly, if the total energy is less than the potential energy at time 
$t_n+l/v_n$, then the particle cannot penetrate region II and simply 
reflects elastically and there is only a change in the sign of the 
velocity; thus the particle gets trapped in region I and returns to the 
wall at $x=0$. After a
time lapse of $2l/v_n$ it will hit again the oscillating barrier and try
again to cross it. If this time the total energy is greater than the
potential energy, then the particle can cross the barrier region;
otherwise, it bounces once more inside region I, and so on.

Now, once the particle overcomes the barrier, it crosses the region II
without changing its velocity $v_n^{\prime }$, even though the barrier is
oscillating in time. When the particle arrives at the right side of the
barrier at $x=l+b$, then another change in the velocity takes place, but
this time the velocity increases in such a way that the kinetic energy $
E_n^{\prime \prime }$ becomes 
\begin{equation}
E_n^{\prime \prime }=E_n^{\prime }+V_0+V_1f \biggl(t_n+{\frac l{v_n}}+ {%
\frac b{v_n^{\prime }}}\biggr),
\end{equation}
where $E_n^{\prime \prime }$ is the energy in region III. Clearly, the time
that it takes to arrive at the wall located at $x=l+b+L$ is $%
l/v_n+b/v_n^{\prime }+L/v_n^{\prime \prime }$, where $v_n^{\prime \prime }$
is the velocity in region III (see Fig. 1b). After a time $%
t_n+l/v_n+b/v_n^{\prime}+2L/v_n^{\prime\prime}$, the particle returns to the
right side of the barrier after traveling twice the distance $L$ in region III, and enters once again region II. However, in general, the potential barrier has a different height, given by $V_0+V_1f(t_n+l/v_n+b/v_n^{\prime}+2L/v_n^{\prime\prime})$. Therefore, the new kinetic energy $E_n^{\prime\prime\prime}$ inside region II is now 
given by 
\begin{equation}
E_n^{\prime\prime\prime} = E_n^{\prime\prime} - V_0 - V_1f\biggl(t_n+{\frac
{l}{v_n}}+{\frac {b}{v_n^{\prime}}} +{\frac { 2L}{v_n^{\prime\prime}}}\biggr),
\end{equation}
Here, once more, there is the possibility that the total energy in region
III is less than the potential energy at time $t_n+l/v_n+b/v_n^{
\prime}+2L/v_n^{\prime\prime}$. In this case, the particle gets trapped in
region III until it can escape by crossing the barrier region.
                        
Finally, after a time $b/|v_n^{\prime\prime\prime}|$, where $
v_n^{\prime\prime\prime}$ is the velocity in region II (see Fig. 1b), 
the particle arrives
at the left side of the barrier at $x=l$, where the velocity varies once
more depending on the height of the barrier at time
$t_n+l/v_n+b/v_n^{\prime}+2L/v_n^{\prime\prime}+
b/|v_n^{\prime\prime\prime}|$. We will denote
the velocity in region I, after this time, by $v_{n+1}$, because this is
precisely the velocity after the next hit with the wall at $x=0$. The last
part of this journey is covered in a time span of $l/|v_{n+1}|$; after this,
the particle hits the wall at the origin at time $t_{n+1}$ and start again
its trip to the oscillating barrier, and the whole process starts again.

Therefore we arrive at the following map in terms of energy and time: 
\begin{equation}
E_{n+1}=E_n^{\prime \prime \prime }+V_0+V_1f(t_n+T_n)
\end{equation}
and 
\begin{equation}
t_{n+1}=t_n+T_n+\sqrt{\frac m{2}} {\frac l{\sqrt{E_{n+1}}}},
\end{equation}
where $T_n$ is given by 
\begin{equation}
T_n=\sqrt{\frac m{2}}\biggl({\frac l{\sqrt{E_n}}}+ {\frac b{\sqrt{%
E_n^{\prime }}}}+ {\frac{2L}{\sqrt{E_n^{\prime \prime }}}}+ {\frac b{\sqrt{%
E_n^{\prime \prime \prime}}}}\biggr)
\end{equation}
and $E_n^{\prime }$, $E_n^{\prime \prime }$ and $E_n^{\prime \prime \prime }$
are given by eqs. (2-4), respectively.

Furthermore, it can be shown that, for this
map, the Jacobian is exactly one, that is, 
\begin{equation}
J={\frac{\partial (E_{n+1},t_{n+1})}{\partial (E_n,t_n)}}=1.
\end{equation}
This result indicates that this map is an area-preserving one \cite{Lich}.

Let us scale the time using the period $\tau $ of the function $f(t)$. We
define the dimensionless quantities: $\phi _n=(2\pi /\tau )t_n$ and $\Phi
_n=(2\pi /\tau )T_n$. In order to scale the energies we introduce the
dimensionless variables: $e_n=E_n/V_0$, 
$e_n^{\prime }=E_n^{\prime }/V_0$,
$e_n^{\prime \prime }=E_n^{\prime \prime }/V_0$ and
$e_n^{\prime \prime \prime }=E_n^{\prime \prime \prime }/V_0$.
With all this definitions we arrive at the following dimensionless map: 
\begin{equation}
e_{n+1}=e_n^{\prime \prime \prime }+1+rf\bigl(\phi _n+\Phi _n\bigr),
\end{equation}
and 
\begin{equation}
\phi _{n+1}=\phi _n+\Phi _n+{\frac{2\pi M}{\sqrt{e_{n+1}}}},\qquad 
(mod 2\pi )
\end{equation}
where $M=l/(w\tau )$, $r=V_1/V_0$ and $w=(2V_0/m)^{1/2}$. Here, $\Phi _n$ is
given by 
\begin{equation}
\Phi _n=2\pi M\biggl({\frac 1{\sqrt{e_n}}}+{\frac bl}{\frac 1{\sqrt{%
e_n^{\prime }}}}+{\frac{2L}l}{\frac 1{\sqrt{e_n^{\prime \prime }}}}+{\frac bl%
}{\frac 1{\sqrt{e_n^{\prime \prime \prime }}}}\biggr).
\end{equation}
This map, although more complicated, resembles the structure of the Fermi
Map \cite{Lich}.

\section{Numerical results}

Let us now analyze numerically the map obtained above. First of
all, we notice that we have four dimensionless parameters: the width of the
barrier $b/l$ scaled with the length of region I; the length $L/l$ of region III scaled with $l$; the ratio of the amplitude of oscillation of the
barrier scaled with its height $r=V_1/V_0$; and $M=l/(w\tau )$. The
parameter $M$ is the ratio of the time of flight $l/w$ in region I of 
Fig. 1a, with velocity $w$, and the period $\tau$ of oscillation of the barrier. That is, $M$ measures the number of oscillations of the barrier since the particle leaves the wall at $x=0$ until it arrives at the left side of the barrier. On the other hand, we will take the periodic function as: 
$f(\phi _n)=\sin(\phi _n)$.

If we fix the barrier position within the one-dimensional box, and choose a
width, then we are fixing the parameters $b/l$ and $L/l$; the remaining two
parameters $M$ and $r$ will control the type of motion. In what follows, 
we take the symmetric case, $b/l=1$ and $L/l=1$, which corresponds to the
oscillating barrier centered inside the box, and an oscillating amplitude
of $r=0.5$.

In Fig. 2 we show the energy-phase space $(e_n,\phi _n)$ for $M=4.7$, 
using the map given by eqs. (9-11). We plot several orbits that
correspond to different initial conditions. We can clearly see that,
for this system,
we have a phase space with a mixed structure, in which we have
periodic, quasiperiodic and chaotic orbits. Some of the fixed points of 
the map can be seen surrounded by elliptic orbits. We notice a fine 
structure of smaller islands in the chaotic region, as is usually the 
case for other maps \cite{Lich}.

The quantity that we want to analyze in detail is the traversal time in
the barrier region, that is, the time it takes the particle to cross
the region where the barrier is oscillating. We can obtain this quantity
simply as $b/v_n^{\prime }$ or
$b/|v_n^{\prime \prime \prime }|$ (see Fig. 1b).
The structure of this traversal or dwell time depends strongly on the 
type of orbit. Clearly, if we have a
periodic orbit, then this time will take only two possible values, 
since $v_n^{\prime }$ and $v_n^{\prime \prime \prime }$ does not change
with $n$. On the other hand, if the orbit is quasiperiodic, the velocity
can vary in a full range of values. In this case, the traversal
time can vary only in a limited range. However, when we have a chaotic
orbit, the variation can display a very rich structure \cite{Mateos}.

For the bounded problem, where the oscillating barrier is confined within
a box, we can obtain a chaotic dynamics as shown in Fig. 2. However, if
we remove the walls and leave only the oscillating barrier, we end up with 
an open system of the scattering type. In this case, we cannot have
chaotic dynamics, since the particle interacts with the barrier only once.
However, we can study not a single particle, but an ensemble of 
noninteracting particles, each of them with different initial conditions.

In Fig. 3 we show a space-time diagram of trayectories for an ensemble of
incident particles. In this case, and for the rest of the figures, we
take $r=0.5$ and $M=77.7$. I use dimensionless distance $x/l$ and
dimensionless time $t$, which is the time scaled with $l/w$. Since 
$l=b$, then $l/w$ is the time it takes to cross the barrier region 
with a velocity $w=(2V_0/m)^{1/2}$. The barrier is located between 
$x/l=1$ and $x/l=2$, and is indicated by horizontal dashed lines in 
Fig. 3. We take an ensemble of initial conditions in which the initial
velocity is constant and the initial phase
is uniformily distributed. We see from Fig. 3 that only a subset of 
particles in the ensemble can cross the barrier region and that the 
traversal time is different for each particle. This is due to the fact that 
each particle is influenced differently by the time-modulated barrier,
depending on the arrival time. That is, different arrival times mean
different barrier amplitudes. 
  
The traversal time is defined as the time it takes to cross the region 
where the barrier is oscillating, and is given by $b/v_n^{\prime }$.
Since we scale this traversal time with the time $b/w$, the
dimensionless form is given by $1/\sqrt{e_n^{\prime }}$.
For the particles in the ensemble, this time is shown in Fig. 4.
We notice that in many cases the dimensionless 
time $t\sim 1$; however, there are some others cases for which $t\gg 1$. 
These large peaks occur when the arrival time is such that the 
total energy is just above the barrier heigth, and thus the velocity 
inside the barrier region is very small and consequently the traversal
time is very large. We can see a strong variation in the traversal
time, that leads to a broad distribution of times. On the other hand,
since the minimum velocity in the barrier region is zero, then there
is no upper bound for the dwell time, and it can acquire very large
values, as seen in Fig. 4.

The traversal time distribution is depicted in Fig. 5. This normalized
distribution has a long-time tail which is a power law.
In Fig. 6 we show the same distribution in a log-log plot that clearly
shows that this is indeed a distribution with a power-law tail
of the form $p(t)\sim t^{-\alpha}$, with $\alpha \simeq 3$. The
straight (dashed) line in this figure has a slope of $-3$.

Another quantity of interest is the transmission coefficient, defined as
the number of particles that cross the barrier region, divided by the
total number of particles in the ensemble. In Fig. 7 we show this
transmission coefficient as a function of $M$. Remember that 
$M=l/(w\tau )$ and is, therefore, proportional to the frequency of
oscillation of the barrier. We can see in this figure that the
transmission coefficient vary strongly with $M$, in particular for low
frequencies ($M\sim 1$). On the other hand, for higher frequencies
($M\gg 1$), the transmission coefficient tend towards a constant value.
This last result indicates that for $M\gg 1$, the oscillating potential 
barrier acts as an effective potential barrier of average height $V_0$.
  
Finally, in Fig. 8 we show the average traversal time as a function of
$M$. Again we can see strong fluctuations of this quantity. Since the
distribution of traversal times is a power law with an exponent 
$\alpha \simeq 3$, we can expect these large fluctuations; although the 
first moment of the distribution is finite in this case, the second or 
higher moments can diverge, leading to these large fluctuations, as is 
usually the case for L\'evy distributions \cite{Levy}.

\section{Concluding remarks}

In this paper, the dynamics of the classical problem 
of an oscillating rectangular potential 
barrier is analyzed. When the oscillating barrier is located 
within a one-dimensional box, we have a bounded problem and the 
corresponding classical dynamics can have a mixed phase space 
structure comprising periodic, quasiperiodic and chaotic orbits.
For the scattering problem of a single oscillating barrier, 
a distribution of traversal times with a power-law tail is obtained. 
This L\'evy-type distribution of times leads to large 
fluctuations of the average traversal time as a function of the 
frequency of oscillation of the barrier; therefore, it is 
difficult to obtain a characteristic time to the process
of crossing the classical oscillating barrier. These large fluctuations
arise due to the sensitive dependence on initial conditions, 
typical of the dynamics of chaotic systems. In particular, 
for our problem, the quantity that controls the traversal time 
is the time of arrival at the barrier. Thus, we obtain a sensitive
dependence on the time of arrival for the classical case.
The possible role for the tunneling time problem, if any, 
of the sensitive dependence on the time of 
arrival and the difficulty to obtain a characteristic
traversal time in the classical domain, remains to be seen. 

\vfill\eject

\vfill\eject

\section{Figure Captions}

Fig. 1 a) Potential well with a rectangular time-modulated potential barrier of width b. The height of the barrier oscillates 
harmonically between $V_0+V_1$ and $V_0-V_1$. b) Typical orbit in 
phase space, showing a general change in the velocity for one 
iteration of the map (solid line) and a second iteration (dashed line).

\medskip

Fig. 2 Phase space $(e_n,\phi _n)$, for $M=4.7$ and $r=0.5$, 
showing periodic, quasiperiodic and chaotic orbits for different 
initial conditions.

\medskip

Fig. 3 Space-time diagram of trayectories for an ensemble of 
incident particles with the same velocity and different phases. 
In this case $M=77.7$ and $r=0.5$. The horizontal dashed lines 
indicate the barrier region.

\medskip

Fig. 4 Traversal time for an ensemble of incident particles. 
In this case $M=77.7$ and $r=0.5$.

\medskip

Fig. 5 Traversal time distribution for the case $M=77.7$ and 
$r=0.5$.

\medskip

Fig. 6 Log-log plot of the traversal time distribution of Fig. 5, 
clearly showing a power law. The slope of the dashed line is $-3$.

\medskip

Fig. 7 Transmission coefficient as a function of $M$, for $r=0.5$.

\medskip

Fig. 8 Average traversal time as a function of $M$, for $r=0.5$.


\begin{thebibliography}{19}

\bibitem{Reviews}
E. H. Hauge and J. A. St\o vneng, 
{\em Rev. Mod. Phys.\/} {\bf 61} (1989) 917;
V. S. Olkhovsky and E. Recami,
{\em Phys. Rep.\/} {\bf 214} (1992) 339;
R. Landauer and Th. Martin,
{\em Rev. Mod. Phys.\/} {\bf 66} (1994) 217;
See also D. Mugnai, A. Ranfagni and L. S. Schulman, eds.,
{\em Proceedings of the Adriatico Research Conference on
Tunneling and its Implications\/} 
(World Scientific, Singapore, 1997).

\bibitem{Bula82}
M. B\"uttiker and R. Landauer, 
{\em Phys. Rev. Lett.\/} {\bf 49} (1982) 1739.

\bibitem{Bula85}
M. B\"uttiker and R. Landauer, 
{\em Phys. Scr.\/} {\bf 32} (1985) 429.

\bibitem{Soko88}
D. Sokolovski and P. H\"anggi, 
{\em Europhys. Lett.\/} {\bf 7} (1988) 7.

\bibitem{Garcia90}
H. De Raedt, N. Garc\'\i a and J. Huyghebaert, 
{\em Solid State Comm.\/} {\bf 76} (1990) 847.

\bibitem{Leav91}
C. R. Leavens and G. C. Aers, 
{\em Solid State Comm.\/} {\bf 78} (1991) 1015.

\bibitem{Leav93}
C. R. Leavens and G. C. Aers,
Bohm Trayectories and the Tunneling Time Problem,
in: R. Wiesendanger and H.-J. G\"untherodt, eds.,
{\em Scanning Tunneling Microscopy III\/}
(Springer-Verlag, Berlin, 1993) 105-140. 

\bibitem{Pimp91}
A. Pimpale, S. Holloway and R. J. Smith, 
{\em J. Phys. A\/} {\bf 24} (1991) 3533.

\bibitem{Azbel}
M. Ya. Azbel, {\em Phys. Rev. Lett.\/} {\bf 68} (1992) 98; 
{\em Phys. Rev. B\/} {\bf 43} (1991) 6847; 
{\em Europhys. Lett.\/} {\bf 18} (1992) 537.

\bibitem{AzMa93}
M. Ya. Azbel and B. A. Malomed, 
{\em Phys. Rev. Lett.\/} {\bf 71} (1993) 1617.

\bibitem{LanMar93}
Th. Martin and R. Landauer, 
{\em Phys. Rev. A\/} {\bf 47} (1993) 2023.

\bibitem{Stov93}
J. A. St\o vneng and A.-P. Jauho, 
{\em Phys. Rev. B\/} {\bf 47} (1993) 10446.

\bibitem{MarSa95}
Ph. A. Martin and M. Sassoli de Bianchi, 
{\em J. Phys. A\/} {\bf 28} (1995) 2403.

\bibitem{Wag95}
M. Wagner, {\em Phys. Rev. A\/} {\bf 51} (1995) 798.

\bibitem{Mateos}
J. L. Mateos and J. V. Jos\'e, 
{\em Physica A\/} {\bf 257} (1998) 434.

\bibitem{Holl}
P. R. Holland, {\em The Quantum Theory of Motion\/}
(Cambridge University Press, 1993).

\bibitem{Leav95}
C. R. Leavens, {\em Found. Phys.\/} {\bf 25} (1995) 229;
{\em Phys. Lett. A\/} {\bf 197} (1995) 88, and references therein.

\bibitem{Lich}
A. J. Lichtenberg and M. A. Lieberman, {\em Regular and
Chaotic Dynamics}. Second Edition. (Springer-Verlag, New York, 1992).

\bibitem{Levy}
See, for instance, M. Shlesinger, G. Zaslavsky and U. Frish, eds., 
{\em L\'evy Flights and Related Topics in Physics\/} 
(Springer-Verlag, Berlin, 1995).

\end{thebibliography}
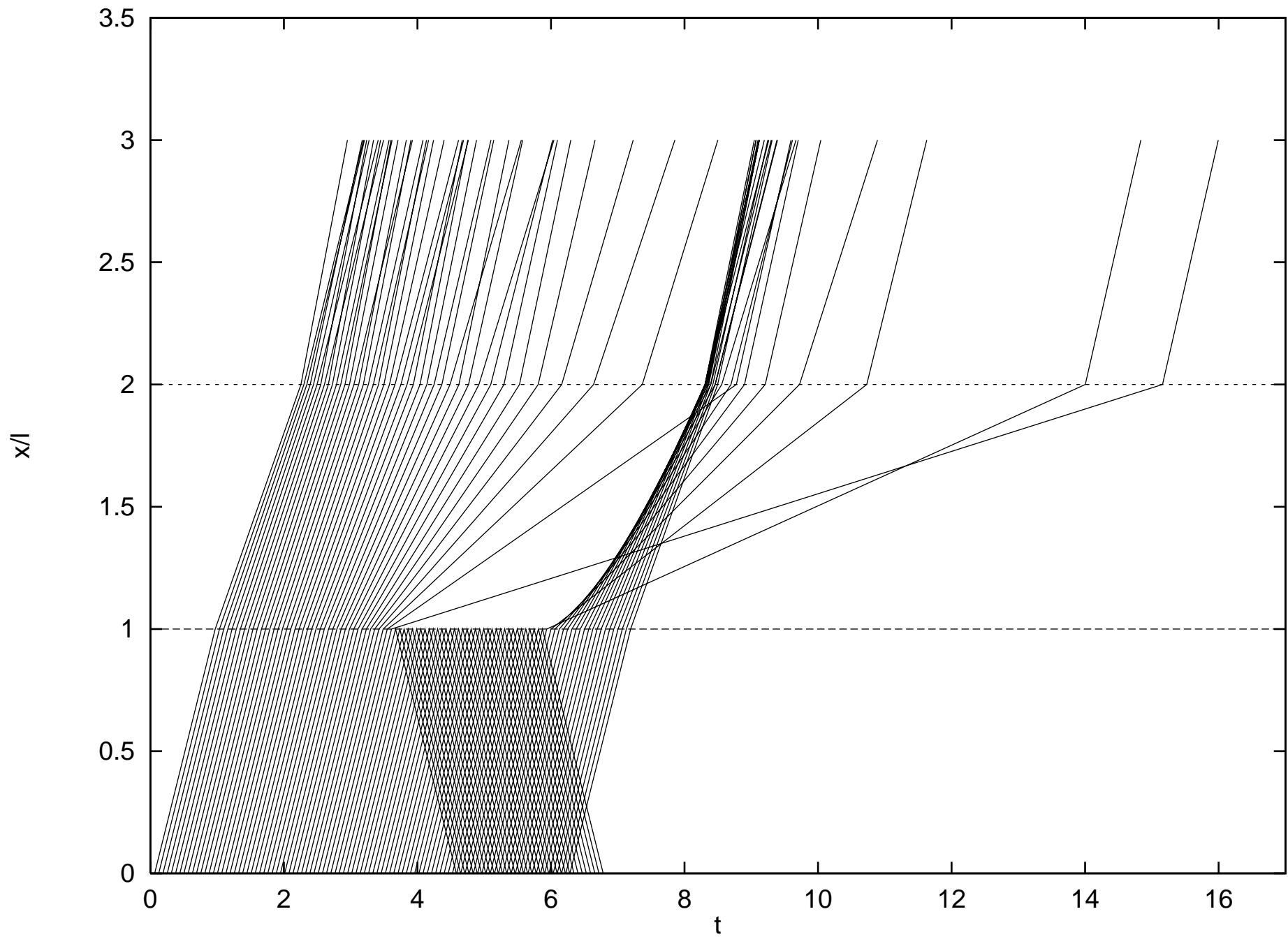
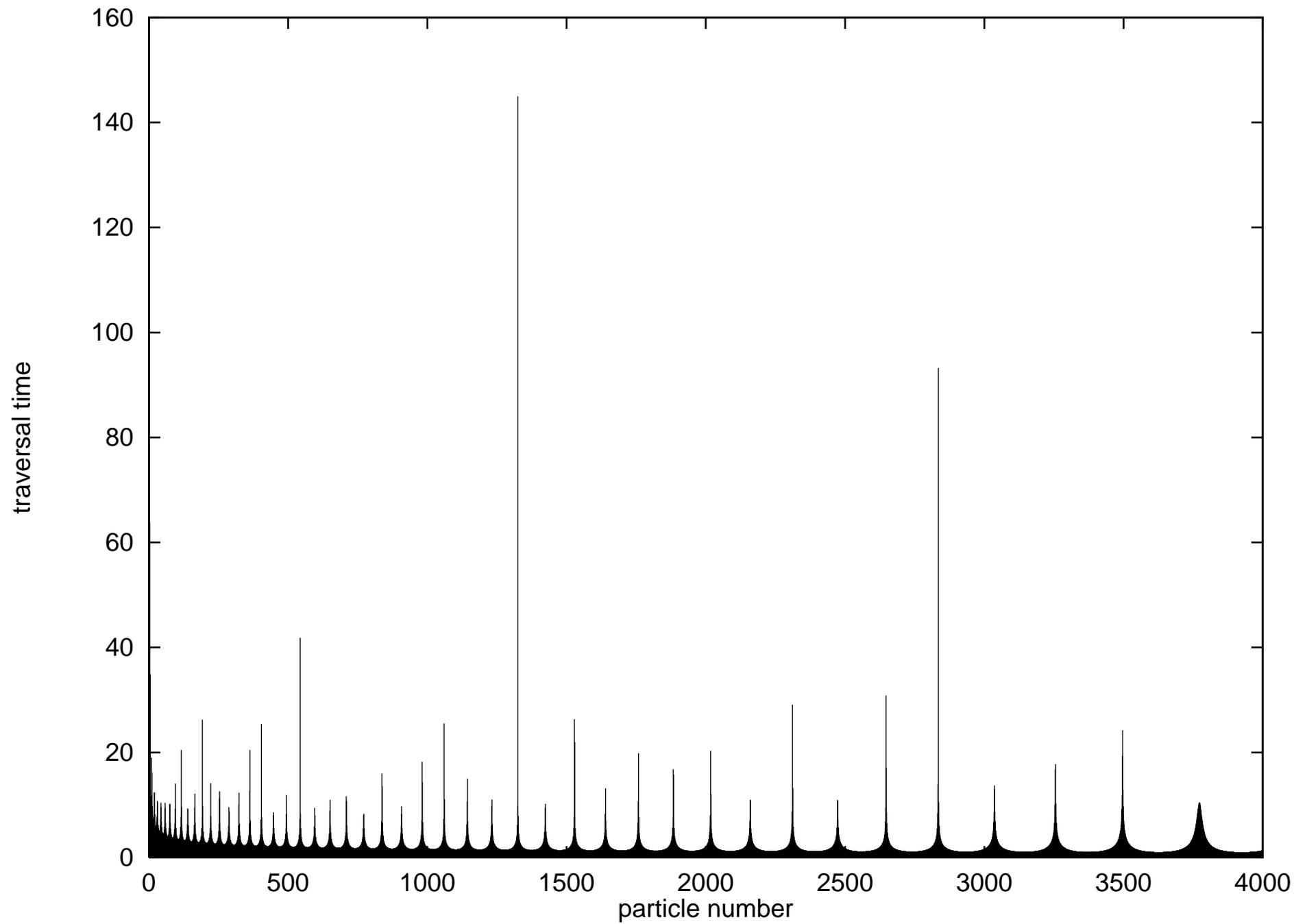
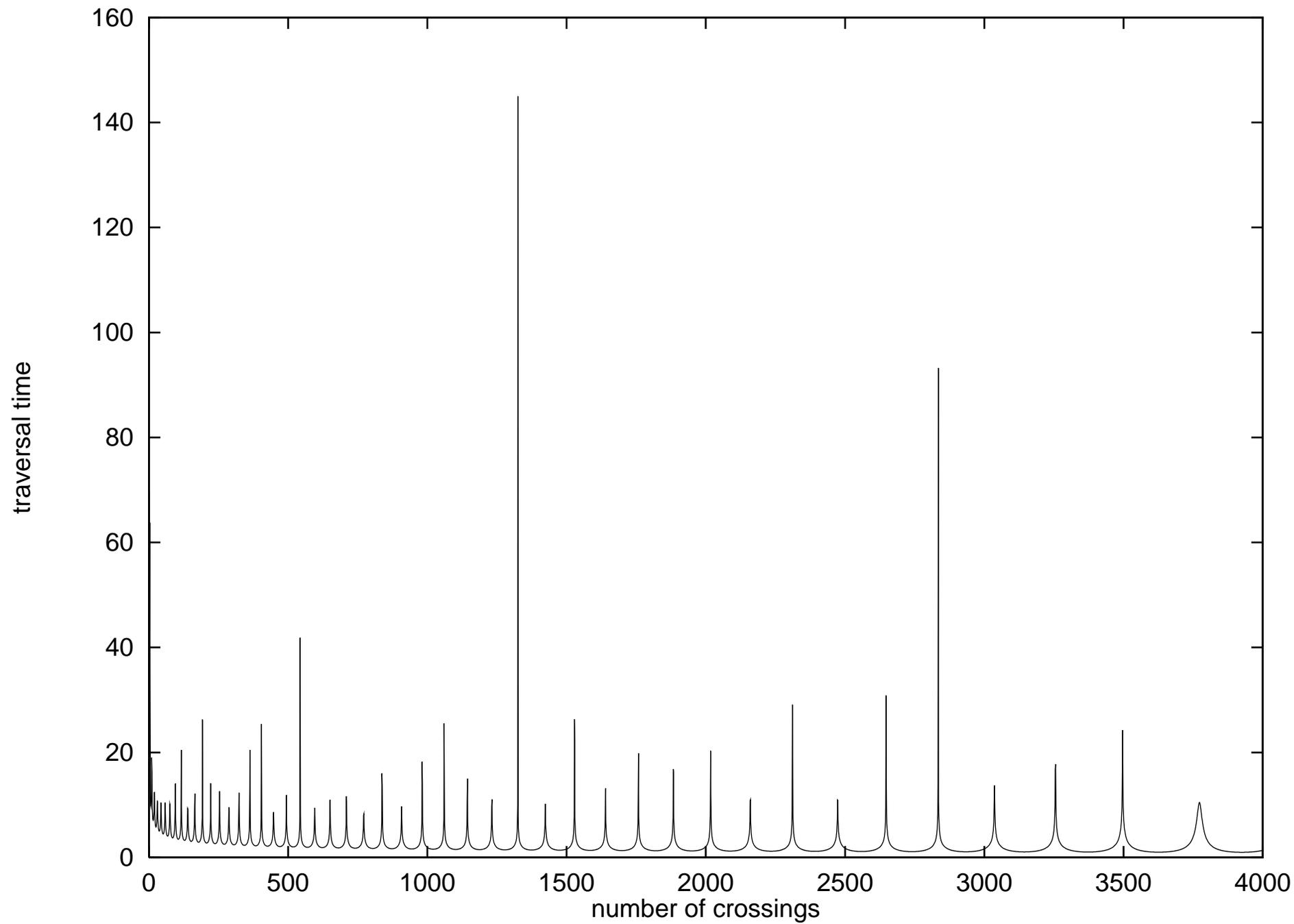
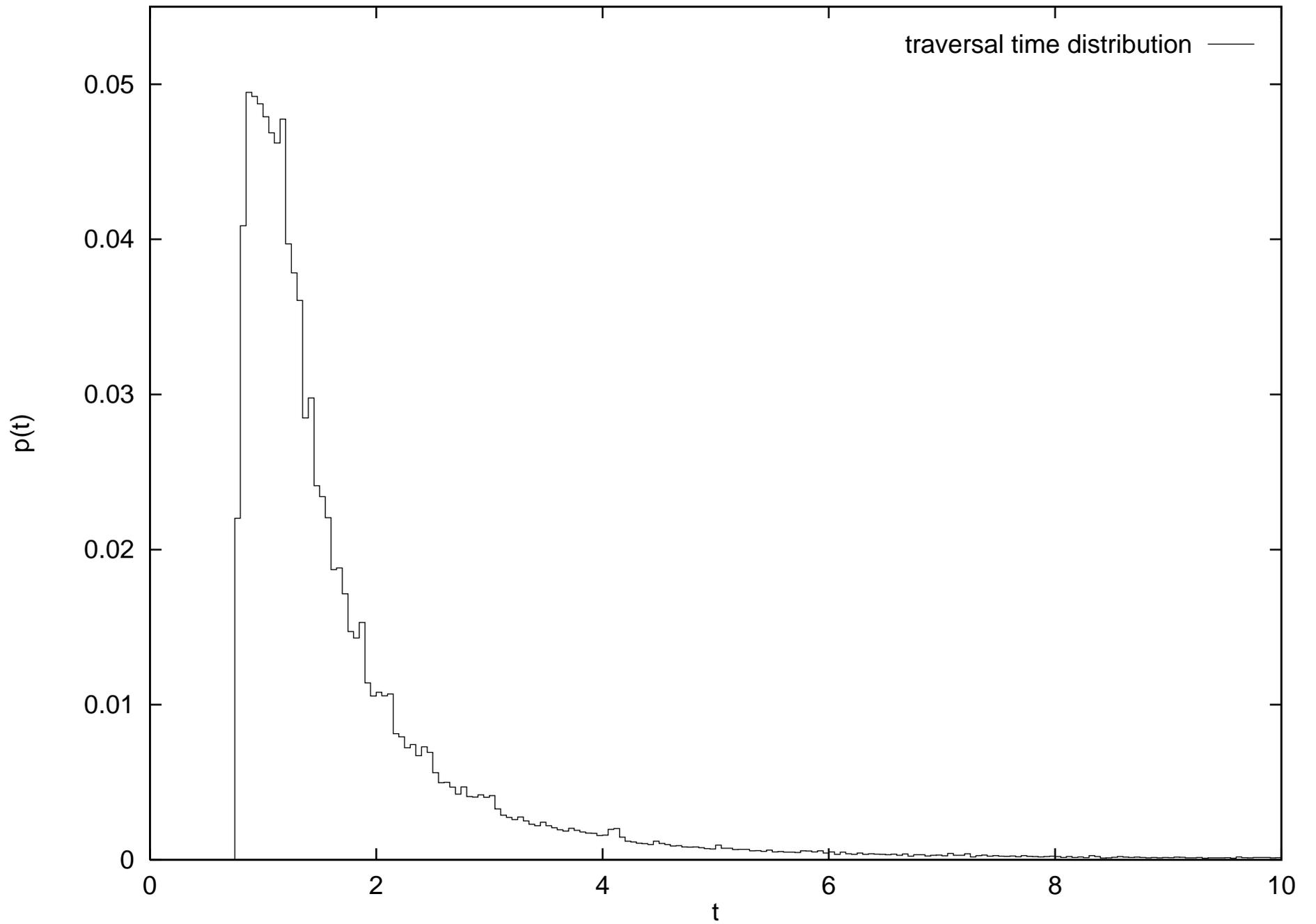
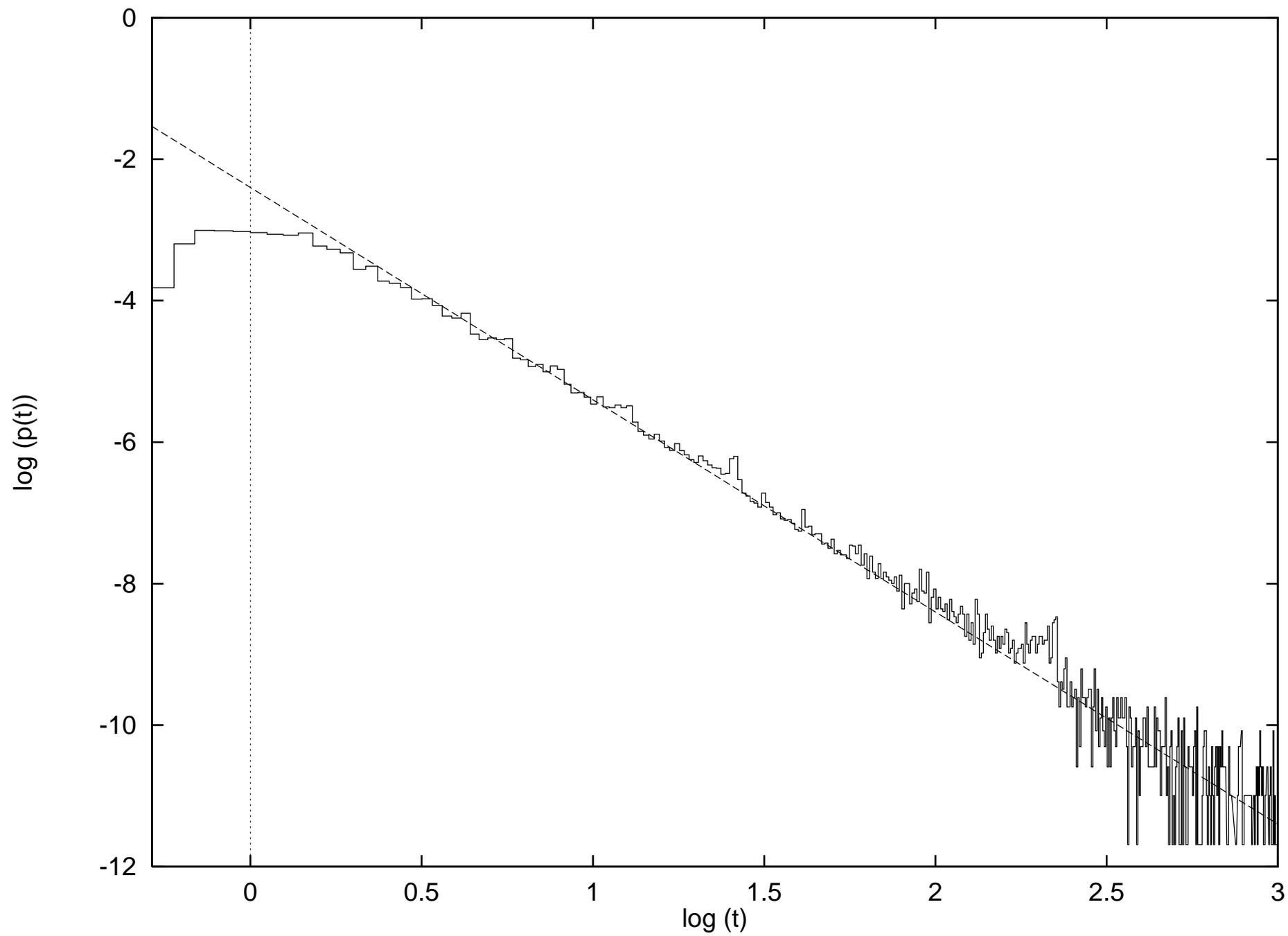
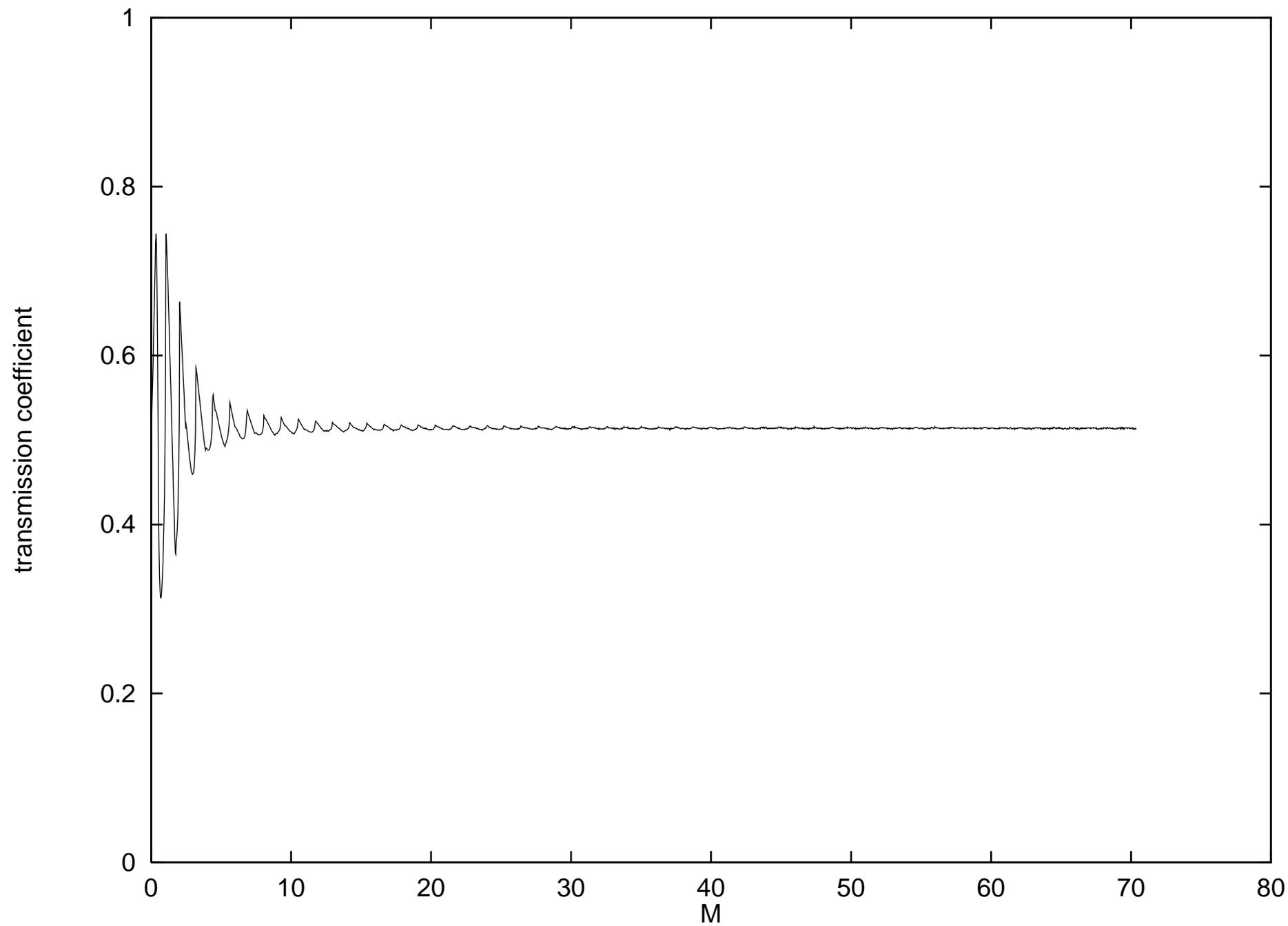
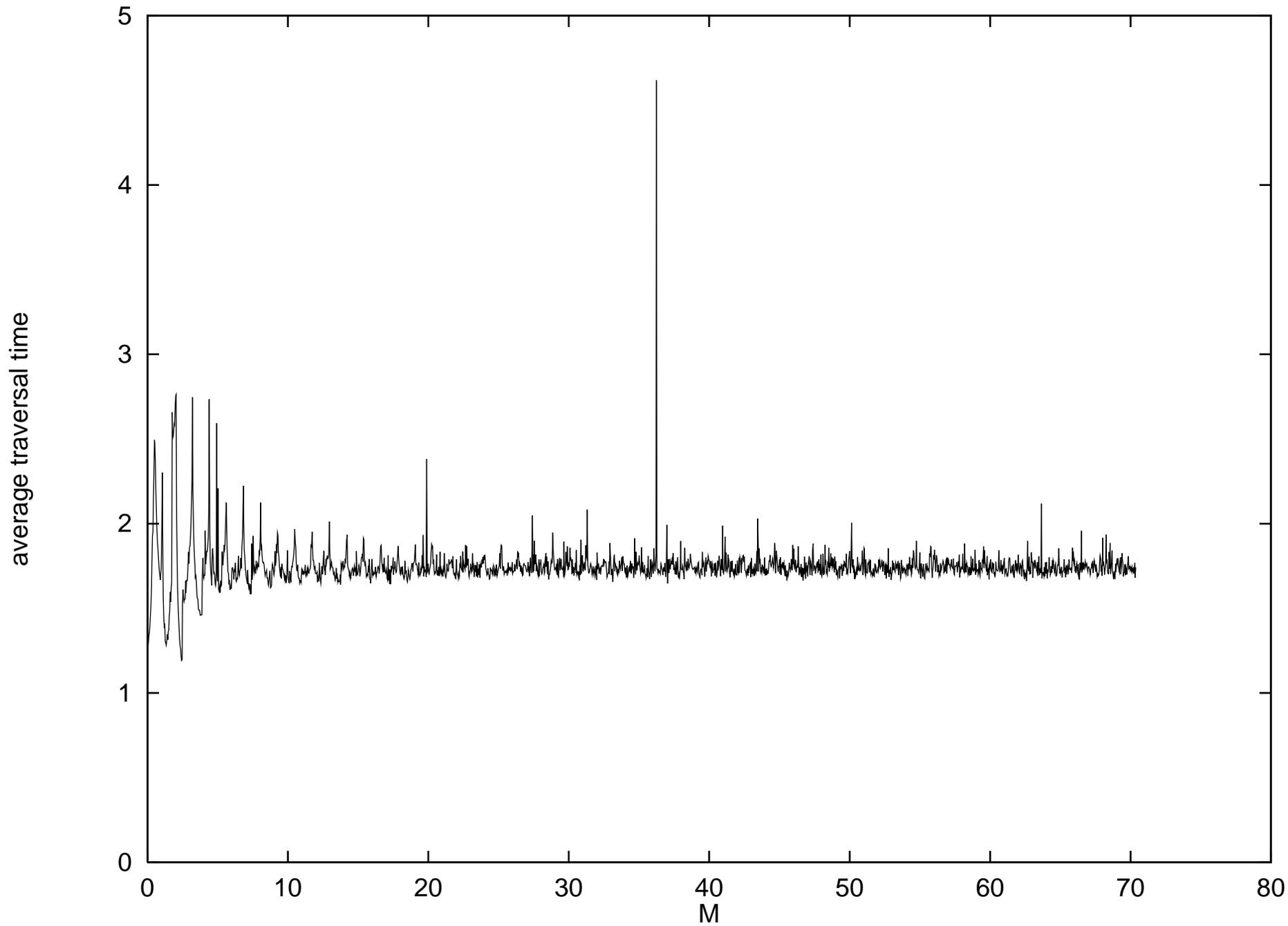
\end{document}